**MAIN POINTS OF THE ARTICLE**

- Non-small cell lung cancer (NSCLC) is a leading cause of mortality. Many treatments for NSCLC rely on combination therapy with bevacizumab, an anti-VEGF adjunct.

- Evidence is presented that scheduling bevacizumab several hours before combination antiproliferatives improves treatment efficacy, via a mathematical model built using patient data collated from 11 clinical trials in NSCLC.

- This work indicates that future trials in NSCLC with bevacizumab should test staggered sequential scheduling. Furthermore, a model fit on a large human clinical trial dataset is provided for future modeling and simulation work in NSCLC.

# OPTIMIZING FIRST-LINE THERAPEUTICS IN NON-SMALL CELL LUNG CANCER: INSIGHTS FROM JOINT MODELING AND LARGE-SCALE DATA ANALYSIS

Authors and Affiliations

Benjamin K. Schneider[1], Sebastien Benzekry[1,2**] and Jonathan P. Mochel[1,3**]

[1]Iowa State University College of Veterinary Medicine, Ames, IA, U.S.A, [2] COMPutational pharmacology and clinical Oncology Department, Inria Sophia Antipolis – Méditerranée, Cancer Research Center of Marseille, Inserm UMR1068, CNRS UMR7258, Aix Marseille University UM105, Marseille, France 3 Precision One Health Initiative, Systems Pharmacology, University of Georgia, Athens (USA)

[**]: co-senior authors

## ABSTRACT

Non-small cell lung cancer (NSCLC) is often intrinsically resistant to several first- and second-line therapeutics and can rapidly acquire further resistance after a patient begins receiving treatment. Treatment outcomes are therefore significantly impacted by the optimization of therapeutic scheduling. Previous preclinical research has suggested scheduling bevacizumab in sequence with combination antiproliferatives could significantly improve clinical outcomes. Mathematical modeling is a well-suited tool for investigating this proposed scheduling modification. To address this critical need, individual patient tumor data from 11 clinical trials in NSCLC has been collated and used to develop a semi-mechanistic model of NSCLC growth and response to the various therapeutics represented in those trials. Precise estimates of clinical parameters fundamental to cancer modeling have been produced—such as the rate of acquired resistance to various pharmaceuticals, the relationship between drug concentration and cancer cell death, as well as the fine temporal dynamics of vascular remodeling in response to bevacizumab. In a reserved portion of the dataset, this model was used to predict the efficacy of individual treatment time courses with a mean error rate of 59.7% after a single tumor measurement and 11.7% after three successive tumor measurements. A delay of 9.6 hours between pemetrexed-cisplatin and bevacizumab administration is predicted to optimize the benefit of sequential administration. At this gap, approximately 93.5% of simulated patients benefited from a gap in sequential administration compared with concomitant administration. Of those simulated patients, the mean improvement in tumor reduction was 20.7%. This result suggests that scheduling a modest gap between the administration of bevacizumab and its partner antiproliferatives could meaningfully improve patient outcomes in NSCLC.

## INTRODUCTION

Lung cancer is the leading cause of cancer mortality nationwide, with an estimated 125,070 deaths expected in the United States this year, 2024 (1). Approximately 85% of those deaths are attributable to non-small cell lung cancer (NSCLC). The need for pharmacological intervention is dependent on the disease stage at the time of diagnosis. If NSCLC is detected early, surgical resection followed by limited adjutant therapy is generally successful in achieving complete remission (2,3). Late-stage NSCLC metastases are often too diffusely spread for surgery or radiotherapy to be effective; therefore, the occurrence of a diagnosis at an advanced stage necessitates treatment strategies predominantly centered on pharmacological interventions. The five-year survival rate for patients diagnosed with localized NSCLC is approximately 58%. However, most patients with NSCLC are first diagnosed at the late stage of the disease, with a five-year survival rate estimated at 6% in the presence of distant metastases (4,5). Ultimately, most patients with mid-to-late-stage NSCLC will be prescribed a platinum-based combination chemotherapy and/or immunotherapy, e.g. bevacizumab-carboplatin-paclitaxel, pembrolizumab-carboplatin-pemetrexed (6–8). The most cutting-edge first-line combination therapies in NSCLC are atezolizumab-bevacizumab-carboplatin-paclitaxel combination therapy (established in IMpower150 study) (9) and pembrolizumab plus pemetrexed-platinum combination therapy (established in KEYNOTE-189) (10,11). Patients with advanced cases of NSCLC are often cycled through several antiproliferative regimens due to high rates of acquired and intrinsic multi-drug resistance (12–15).

Acquired or intrinsic drug resistance is a major cause of first-line therapeutic failure in NSCLC. In a previous study of resected NSCLC by d'Amato et al., intermediate to extreme resistance to carboplatin was found in 68% of NSCLC cultures (vs. 63% and 40% for cisplatin and paclitaxel, respectively) (12). Likewise, in the recent KEYNOTE-001 study, NSCLC patients receiving pembrolizumab had an objective response rate of 19.4%, indicating that a vast majority of individuals did not significantly respond to therapy (15). Finally, in the most recent ARIES observational cohort study of first-line NSCLC treatment involving bevacizumab, the failure rate for 1967 patients was approximately 51% (16). Taken together, there is a *clear and <u>critical unmet clinical need</u>* to improve patient response to treatment and clinical outcome in NSCLC.

Improving clinical outcome in NSCLC with existing therapeutics requires addressing at least three prominent challenges: **1).** The disease is often late stage at the time of diagnosis, and treatment relies heavily on complex combination therapies. However, oncologists currently have few evidence-based tools to individualize and optimize scheduling of those therapies; **2).** The success rate of first-line therapy is typically low, therapeutic dosing (amount, interval, gap) strategies are often unoptimized, and resistance to treatment is rapidly acquired (6–8). **3).** The window between the threshold for therapeutic efficacy and drug overdose can be narrow, especially with the relatively long elimination half-life of most antiproliferatives. Meeting these challenges are made even more difficult by the relatively sparse and sporadic collection of pharmacokinetic and pharmacodynamic data in clinical studies involving cancer.

The status quo as it pertains to therapeutic management and dosing decisions in NSCLC is that medications are administered according to recommended guidelines, and those guidelines are in turn established through one-to-one human clinical trials. A major limitation in this approach is that it mainly focuses on the *average* patient, such that clinicians have little evidence-based guidance to help them *individualize* treatment in patients who are unresponsive or achieve partial response to therapy.

Nonlinear mixed-effects (NLME) population modeling is a highly relevant computational framework for exploring optimization of drug dosing schedules on an individual level, without the considerable time and resource investment required to conduct a suite of one-to-one clinical trials. With enough data, NLME modeling has the potential to overcome some of the intrinsic limitations of antiproliferative optimization through simulation of alternative schedules and clinical trials. And though measurements in any given NSCLC patient may be sparse, if enough high-quality data can be pooled between studies a precise predictive model can be built from the richness of the variation between clinical trials.

NLME models are designed to continuously incorporate new information, and can easily be expanded by incorporating new data to simulate relevant experimental designs and guide treatment decisions, such as: *What happens if dosages are halved, but administrations are made twice as often? If some pre-existing condition requires scheduling modification (e.g., renal damage and cisplatin nephrotoxicity), how should dosages of concurrent chemotherapeutics be modified? What is the ideal time after first dose to start looking for signs of acquired resistance?*

There are several paradigms for the development of NLME models. Some models, called quantitative systems pharmacology models, attempt to feature every element of the biological system in mathematical detail (17,18). Others are purely empirical and have the barest of structures. In this study, a semi-mechanistic modeling approach has been adopted to balance any limitations in the datasets with precise mathematical descriptions of NSCLC pathology developed in many previous studies. Semi-mechanistic models attempt to balance the desire for biologically relevant mechanisms with limitations in the datasets by modeling any prominent pathology with generalized empirical formulations.

Prior to this study, a preliminary semi-mechanistic model of NSCLC growth dynamics was developed in a preclinical experimental setting (19). Scaling this model to human biology, it was predicted that scheduling bevacizumab administration 1.2 days before pemetrexed-cisplatin administration has the potential to significantly improve tumor reduction (>50%) as opposed to concomitant dosing (20). In this follow-up study, the aim is to develop a predictive mathematical framework of first- and second-line therapeutic drugs for NSCLC by iterating on previous modeling efforts using a large (~11,000 patients) clinical trial dataset made available through the *Vivli.org* database (**Table 1**) (21).

The overall objectives in the present analysis were to build a model which **1).** generalizes a previous model of NSCLC growth and response to bevacizumab-pemetrexed-cisplatin to the greater set of combination therapies and modes of action represented in this large clinical database, and **2).** characterizes the time-course of resistance for those currently registered therapeutics for NSCLC.

In building this model, several empirical approximations of clinical parameters *fundamental* to cancer modeling have been estimated such as the rate of acquired resistance to various pharmaceuticals, the relationship between drug concentration and cancer cell death, as well as the fine temporal dynamics of anti-VEGF therapy. Then, using this high-powered model, an up-to-date prediction of optimal bevacizumab-pemetrexed-cisplatin was made, in patients with NSCLC.

## MATERIALS AND METHODS

*Literature Search and Review*

To build this model, access to data was sought using very broad criteria. Data was prioritized from clinical trials where bevacizumab had been used in combination with other therapeutics to treat NSCLC. To build on previous modeling efforts, it was necessary for studies to include records of *individual patient tumor sizes* over time. After reviewing several potential data custodians to partner with, an application was sent for data access through the Clinical Study Data Request (CSDR) portal. Eleven different studies were identified which were available through CSDR's platform which met this study's requirements requirements (**Table 1**) and access was permitted to a secure server containing the datasets beginning on January 6th, 2020.

Due to contractual obligations between the data administrators (Roche, Eli Lilly) and data access provider (CSDR), the project files were moved to a new secure server managed by Vivli on September 22nd, 2020. During this transfer process, access to five additional datasets was requested involving newer standard-of-care therapeutics and NSCLC, but these data were not included in the final modeling dataset.

*Data Processing and Collation*

Despite efforts by organizations like The Clinical Data Interchange Standards Consortium (CDISC) to standardize dataset creation, internal standards for data annotation, units of measurement, file structure, and abbreviated shorthand are highly variable between datasets (22). This variability makes collating and preparing large datasets for mathematical modeling a challenging and highly error-prone task.

To reduce the possibility of introducing mistakes into the datasets during collation, iterative collation of datasets was systemized. In practice, this meant producing a single summary dataset from each study before collating these summary datasets into a comma-delimited values file (i.e. .csv). For each study, the data record files containing the most relevant information to the study were first identified. The files of primary interest contained anonymized individual identifiers, tumor identifiers, tumor pathology, tumor size, and drug administration details all organized longitudinally. Data from these files were imported directly into R (version 3.5.2) to be normalized for import into the NLME parameter estimation software (Monolix Suite version 2020R1, Lixoft). Normalization consisted of formatting the data as recommend by Lixoft (15), as well as matching units to a dictionary of standards maintained throughout the study.

After each dataset was produced, the data was explored visually with a combination of R 3.5.2 as well as Datxplore software (Monolix Suite version 2020R1, Lixoft) to check for potential outliers and errors. Plots produced from the raw imported data were compared with plots produced from the transformed data in R 3.5.2 to improve quality and consistency in processing. Finally, the individual datasets produced in analyzing each study were bound into one single comma separated value file. As a final quality check, the final dataset was again re-imported into R 3.5.2, and subset down to individual studies. Then, the processed studies were again compared with the raw files from the clinical trials for consistency. Data were both received and maintained in a fully anonymized format to protect patient privacy.

*Non-Linear Mixed Effects Modeling and Characterizing Individual Variation*

The recorded data ($y_{ij}$) were pooled and used to estimate model parameters via the stochastic approximation expectation maximization algorithm (SAEM) implemented in Monolix (23). After estimating population parameters ($\boldsymbol{\mu}$) and variance, individual parameters ($\boldsymbol{\phi}_i$) were estimated using the modes of the individual estimated posterior distributions. The posterior distributions were estimated using a Markov-Chain Monte-Carlo (MCMC) procedure. NLME models were written as previously described (Sheiner & Ludden, 1992) (**Eq. 1**).

**Equation 1:**

$$y_{ij} = F\left(\phi_i, \beta_i, t_{ij}\right) + G\left(\phi_i, t_{ij}\right) \cdot \varepsilon_{ij} \vee \varepsilon_{ij}\ N(0, \sigma^2)$$

$$\phi_i = h(\mu, \eta_i, \beta_i) \vee \eta_{ij}\ N(0, \Omega, \omega^2)$$

$$j \in 1, \ldots, n_i, i \in 1, \ldots, N$$

Briefly, model predictions ($F(\boldsymbol{\phi}_i, \boldsymbol{\beta}_i, t_{ij})$) for the $i^{th}$ individual at the $j^{th}$ timepoint were written as a function of individual parameters, individual covariates ($\beta_i$), and time ($t_{ij}$). The residuals were modeled as $G(\boldsymbol{\phi}_i, t_{ij}) \cdot \boldsymbol{\varepsilon}_{ij}$. Individual parameters were modeled with function $h(\boldsymbol{\mu}, \boldsymbol{\eta}_i, \boldsymbol{\beta}_i)$. Interindividual variability, $\boldsymbol{\eta}_i$, are distributed normally with mean **0**, variance-covariance matrix $\boldsymbol{\Omega}$, and variance $\omega^2$. Typically, $h(\boldsymbol{\mu}, \boldsymbol{\eta}_i, \boldsymbol{\beta}_i)$ is a lognormal link function (**Eq. 2).** In cases where $\phi_i$ is bounded, $h(\boldsymbol{\mu}, \boldsymbol{\eta}_i, \boldsymbol{\beta}_i)$ was typically a logitnormal link function (**Eq. 3**).

**Equation 2:**

$$\phi_i = \mu \cdot e^{\eta_i + \beta_i}$$

**Equation 3:**

$$log\left(\frac{\phi_i}{1-\phi_i}\right) = log\left(\frac{\mu}{1-\mu}\right) + \eta_i + \beta_i$$

*Model Building*

The key focal points of this project encompassed the precise modeling of individual tumor growth and response, the formulation of a mathematical framework that accounts for both acquired and intrinsic resistance, the characterization of individual drug effects and interactions, and the modeling of the temporal dynamics underlying tumor vascularization improvement through bevacizumab dosing.

Candidate models were built in multiple development phases. First, previously established models of drug pharmacokinetics (PK) were reproduced within the modeling project using the Mlxtran language (Monolix Suite version 2020R1, Lixoft) (24). Then, several sample datasets were created, with between 5% and 10% of the complete dataset for initial model building. This shortened calculation time and reduced computational complexity during model building. Next, several base candidate models were chosen for NSCLC pharmacodynamics (PD) and were finally implemented using the Mlxtran language.

Two well-established models of NSCLC were used as the base candidate structure of tumor growth and drug PD. The first model fit to the data was the Claret model of tumor growth and response to antiproliferatives (25). The second model fit was built on a previous Gompertzian model of BEV-PEM/CIS (20,26). Initially, all antiproliferatives were assumed to operate in a simple log-kill manner (27).

Finally, using a combination of manual exploration and modified SAEM searches with simulated annealing on the $\boldsymbol{\eta}_i$ terms, several reasonable parameter estimates with which to initialize the parameter search were identified. Lastly, the best performing candidate models were finalized and compared for performance using quality-of -fit and robustness of parameter estimates, as previously described (28–31). Any modifications to structure would be iterated on using a similar method and compared using a suite of model evaluation techniques.

*Model Evaluation*

**Quality of fit** was determined using both goodness-of-fit plots and summary statistics. The stability of parameter estimates was assessed through a combination of methods, including examination of the SAEM search, verification of the attainment of auto-stop criteria implemented in Monolix, and validation of consistent convergence to the same set of parameter estimates when using randomized initial starting parameters within a local range.

**Accuracy of individual fits** was assessed using a sample of individual plots, an observations-vs-predictions plot using the full conditional distribution, a scatter plot of the residuals, as well as the corrected Bayesian information criteria (BIC) – estimated via importance sampling.

**Assumptions of variance** were validated by plotting the conditional distribution against theoretical distributions as well as standard statistical tests. Within the NLME framework, random effects and residuals are assumed to be predictably distributed – usually normally or functionally-linked to normal. The Shapiro-Wilk and Van Der Waerden tests were used to determine normality and symmetry of distributions around the mean, respectively. Correlation between individual parameters and either other individual parameters or covariates were tested with a Pearson's correlation test or ANOVA, respectively. Plotting these relationships assisted in determining the nature of these correlations.

**Precision and accuracy** of the final model was assessed to evaluate models against one another. Precision of parameter estimates was made using relative standard error (RSE). Overall model quality (accuracy) of fit was evaluated using the corrected BIC (BICc). Diagnostic plots assisted in comparing models with similar overall performance.

**External validation and overfit assessment** for the final model were made using individualized Bayesian predictions. In short, only 85% of the data was used for model building. After the final model was fit, all population parameter estimates were fixed to their typical value and the model was then fit to the reserved 15% of the data. The goodness-of-fit of the model to the external dataset was used as both an external validation tool and for assessing model overfit.

*Clinical Trial Simulations*

The objective of the clinical trial simulations was to assess the potential advantages of sequential administration of pemetrexed-cisplatin therapy with bevacizumab. For this purpose, a cohort of 5,000 virtual individuals was simulated based on the final fit of the model, which was subsequently replicated across 31 different sets of test conditions for sequential administration of bevacizumab and chemotherapy. These test conditions were derived from modifications of the Multicenter Phase II MAP Study of NSCLC (22):

- Pemetrexed administered every 22 days for 7 cycles at 945 mg and with an infusion time of 10 minutes (an average between the administration schedules in the datasets, although not aligned with typical standard of care).
- Cisplatin was administered every 22 days for 3 cycles at 135 mg and with an infusion time of 30 minutes.
- Bevacizumab was administered every 22 days for 7 cycles at 795 mg and with an infusion time of 30 minutes.

The administration of bevacizumab followed the administration of pemetrexed and cisplatin with a time gap ranging from 0 to 3 days, in increments of 0.1 days. This approach aimed to evaluate whether concurrent administration yielded superior outcomes compared to administering bevacizumab after a delayed interval following pemetrexed and cisplatin. All simulations were executed using Simulx (Monolix Suite version 2023R1, Lixoft). Subsequently, the most effective time gap was identified among all the gaps tested within the range of 0 to 3 days.

## RESULTS

*Data Summary*

Data were received in directories of fully anonymized data frames (e.g. excel files, SAS files, *.csv, etc.) organized by general category of data. For initial data processing, data were explored for bulk trends, standardized, and data with suspected errors and outliers were removed. Studies were then collated in a single data frame designed for use with the Monolix suite (2020R1). Between the 11 studies, 3,686 patients data were determined to be potentially suitable for analysis. After exploring the datasets in greater detail, 2,586 patients were determined to gather all necessary data to create models of tumor growth and response, including unique patient IDs, a time recorded for each dosing and measurement event, individual tumor diameter measurements, and details of administration (**Table 1**). See Supplemental 1 for further information on inclusion exclusion criteria and the linked publications for further demographic information. Due to the massive effort of this study and semi-mechanistic aims, demographic descriptions of the dataset are limited in this manuscript.

The individual tumor longest diameter time-course was chosen as independent variable in the model, as the more classical sum of the longest diameters (SLD) did not perform as well in semi-mechanistic models of tumor growth in this study. Statistically, all patient-tumor combinations were treated as unique subject-occasions. So, within-subject inter-tumor variability was included, but no distinction was made between within-subject variability and between-subject variability. While acknowledging the potential correlation of individual parameters across multiple tumors within an individual, it is important to note that the approach described above represents a compromise that strikes a favorable balance between modeling systems biology and model parameters structural

identifiability—a chief example of the types of decisions made when constructing semi-mechanistic models. Among the subject-occasions, there were 6,197 unique tumors belonging to 2,586 patients.

After a short period of testing, several further restrictions were imposed on the dataset to facilitate numerical stability of the modeling. The first condition was that tumors were required to have been measured three or more times to qualify for inclusion. This reduced the number of unique tumor IDs to 4,701 and unique individuals to 2,036. If for each sample $y_{ij}$, for individual *i* at time *t*, $y_{ij}$was greater than or equal to $y_{ij-1}$, the tumor was labeled as a monotonic non-responder and excluded from the dataset. This reduced the number of individual tumors to 4,473 and individual IDs to 1,977. After removing these data, there were 4,450 tumors and 1,963 individuals left in the dataset. These restrictions imposed on the initial dataset representing 2,586 patients reduced the number of samples from 29,885 to 26,515. This is an approximate 11% reduction in data. The reduction in the dataset size was primarily due to the limited frequency of tumor size assessments among the majority of patients in the clinical studies. Lastly, approximately 85% of the available data from each study (randomly allocated by subject-tumor pairs) was used for model building and the remaining 15% of data was reserved for external validation of the final model.

*Model Building*

Individual pharmacokinetic parameters were estimated in relatively few of the clinical trials referenced herein. Therefore, population pharmacokinetic models were collated from scientific literature (**Table 2**). Between-patient variability in drug pharmacokinetics was not included in the pharmacokinetic model as it led to practical unidentifiability of random effects parameters.

For pharmacodynamic modeling, two primary candidates were evaluated for the base description of tumor growth and response; **1).** the **Gompertzian model** of tumor growth and **2).** the **Claret model** of tumor growth (25,27). Both models describe tumor volumes, where the raw data reported tumor longest diameter. The tumor diameters were converted into volumes by making the assumption that the tumors were approximately spherical in shape. Using the Bayesian information criteria as a parsimonious method of cross-evaluating models, the Gompertzian model outperformed the Claret model of tumor growth during several stages of structural model development.

In the Gompertzian model of tumor growth (**Eq. 4a**), the unperturbed tumor grows at a rate $\alpha$ and is exponentially limited in growth by parameter $\beta$. $v_c$ is a scaling factor relating individual tumor cell turnover to volume (26). It was set to $10^6$ cells/mm$^3$ which is the classical assumption of approximate number of cells per unit volume (14).

**Equation 4**

**(4a)**

$$\frac{dv}{dt} = \left(\alpha \cdot Q_\alpha - \beta \cdot log\left(\frac{v}{v_c}\right)\right) \cdot v - Q_\gamma \cdot v$$

$$log(Q_\alpha) = -\left(1 + w_{bev_\alpha} \cdot cc_{bev}(t-\tau)\right) \cdot \left(eff_{microt} + eff_{vegf}\right)$$

$$Q_\gamma = \left(1 + w_{bev_\gamma} \cdot cc_{bev}(t-\tau)\right) \cdot \left(eff_{plat} + eff_{afolate} + eff_{dr5} + eff_{egfr} + eff_{dnasub}\right)$$

**(4b)**

$$\frac{dv}{dt} = \left(\alpha - \beta \cdot log\left(\frac{v}{v_0}\right)\right) \cdot v - Q_\gamma \cdot v - Q_\rho \cdot v + kk_{inj} \cdot v_{inj} \cdot p$$

$$\frac{dv_{inj}}{dt} = Q_\rho \cdot v - kk_{inj} \cdot v_{inj} - Q_\gamma \cdot v_{inj}$$

$$Q_\delta = \left(1 + w_{bev_\delta} \cdot cc_{bev}(t - \tau)\right) \cdot \left(eff_{microt} + eff_{vegf}\right)$$

$$Q_\gamma = \left(1 + w_{bev_\gamma} \cdot cc_{bev}(t - \tau)\right) \cdot \left(eff_{plat} + eff_{afolate} + eff_{dr5} + eff_{egfr} + eff_{dnasub}\right)$$

In **Equation 4a**, $Q_\alpha$ and $Q_\gamma$ are the antiproliferative effects resulting in growth reduction and irreversible cell death, respectively. Chemotherapeutics which acted on the microtubules – docetaxel and paclitaxel – along with the direct effect of bevacizumab were included in $Q_\alpha$. All other therapeutics were modeled as drugs resulting in irreversible cell death. Transient enhancement in efficacy via tumor vasculature normalization by bevacizumab was modeled as occurring at time $(t - \tau)$ to account for the time delay between administration and efficacy enhancement i.e., $\tau$ (19).

**Equation 4a** heavily exaggerated the effect of bevacizumab in limiting cell growth rates. To account for this, a second compartment was implemented which represented reversible cell injury, $v_{inj}$, from which cells could return to the primary volume (i.e. $v$) from (**Eq. 4b**). In this equation, elimination rate from tumor cell injury is governed by intercompartmental transfer rate $kk_{inj}$ as well as proportion of repaired cells returned to the unperturbed cycle of proliferation, $p$ (value fixed between 0 and 1 using logit-link). Noteworthily, the scaling factor $v_c$ was set to the initial tumor volume to tie growth limit to initial tumor size. This modified model structure provided a more accurate representation of the growth-limiting effects exerted by bevacizumab, paclitaxel, and docetaxel, striking a balance that avoided the exaggerated effect observed in **Equation 4a**.

Irreversible cell death was modeled as occurring over a series of transitions between several compartments with intercompartmental transfer rate $kk$. The final tumor volume, a summation of the primary tumor volume and death compartments ($z_1$, $z_2$, and $z_3$,) as well as injured cell volume $v_{inj}$, was then transformed to a tumor diameter to match the (observed) independent variable in the dataset (**Eq. 5**).

**Equation 5**

$$\frac{dz_1}{dt} = Q_\gamma \cdot v + Q_\gamma \cdot v_{inj} + kk_{inj} \cdot v_{inj} \cdot (1-p) - kk \cdot z_1$$

$$\frac{dz_2}{dt} = kk \cdot z_1 - kk \cdot z_2$$

$$\frac{dz_3}{dt} = kk \cdot z_2 - kk \cdot z_3$$

$$n = v + z_1 + z_2 + z_3 + v_i$$

$$TumorDiameter = 2 \cdot \left(\frac{3n}{4\pi}\right)^{\frac{1}{3}}$$

For the pharmacodynamic effect ($eff$) of the various therapeutics on tumor growth, a version of a previously published log-kill model was implemented, whereby each drug concentration was scaled by a single parameter (represented by $w_d$ for weighting for drug $d$) and the weighted sum of those concentrations determined the overall cytotoxic effect (**Eq. 6a**). $cc_d(t)$ represents the concentration of drug $d$ at any arbitrary time $t$. This formulation proved slightly unstable, so the sum of effects was eventually grouped under an inverse *logit* function, so that this sum would be bounded to a value comprised between 0 and 1 (**Eq. 6b**). After several more iterations of the model, the fit was further improved by individualizing drug effects relative therapeutic mechanism of action.

Lastly, resistance to anticancer treatment was modeled using a variant of an exposure-response model (32). In this mathematical representation, the tumor cells were assumed to become increasingly more resistant to treatment (rate governed by parameter $\lambda_d$) as a function of drug exposure (area under the concentration time-curve, AUC) – **Eq. 6c**. Additionally, the effect of drugs with similar mechanisms of action were weighted relative to each other—i.e. $w_{d_{1-rel-2}}$ represents the weight of some drug (given the arbitrary index 1) relative to some other drug (drug 2) which shares a mechanism of action, e.g. cisplatin and carboplatin. A complete diagram of the model is available for review in **Figure 1**.

**Equation 6**

**(6a)**

$$eff_d(t) = \left(w_d \cdot cc_d(t)\right)$$

$$eff_{total}(t) = \sum w_d \cdot cc_d(t)$$

$$d \in cis, car, pem, apo, erl, gem, pac, doc, bev$$

**(6b)**

$$eff_d(t) = \left(w_d \cdot cc_d(t)\right)$$

$$eff_{total}(t) = \gamma \cdot \left(invlogit\left(\sum w_d \cdot cc_d(t)\right) - 0.5\right)$$

$$d \in cis, car, pem, apo, erl, gem, pac, doc, bev$$

**(6c)**

$$AUC_d = \int cc_d(t)$$

$$\begin{bmatrix} singledrugeff(t) \equiv & log\left(eff_m(t)\right) = log\left(w_d \cdot cc_d(t)\right) - \left(\lambda_d \cdot AUC_d(t)\right) \\ paireddrugeff(t) \equiv & log\left(eff_m(t)\right) = log\left(w_{d_{1-rel-2}}\left(cc_{d_1}(t) + w_{d_2} \cdot cc_{d_2}(t)\right)\right) - \lambda_{d_{1-rel-2}}\left(AUC_{d_1}(t) + \right. \end{bmatrix}$$

$$d \in cis, car, pem, apo, erl, gem, pac, doc, bev$$

$$m \in plat, afolate, dr5, egfr, dnasub, microt, vegf i.e. modeofaction$$

Individual variability was modeled using the standard lognormal distribution and initial tumor volumes were fixed to the measurement of the tumor at time 0, relative to each individual. The only exception was made for parameter *p* which was fixed between 0 and 1 using a logit-link function (**Eq. 7**). At last, measurement error was modeled using the function *combined 1* in Monolix, i.e., including a single additive term (a) added with a single proportional term (b).

**Equation 7**

$$\text{log-link: } \phi_i = \phi_{pop} \cdot e^{\eta_i}$$

$$\text{logit-link: } logit(p_i) = logit(p_{pop} + \eta_i)$$

$$v(t = 0) = y(t = 0)$$

<u>*Model Diagnostics*</u>

Graphical evaluation of model diagnostics support the validity and predictive performances of the final selected model. Specifically, the SAEM search was stable and reliable when estimating the final set of parameter estimates (**Figure 2**). Individual fits were reasonably descriptive with both a well captured tumor growth inhibition in response to treatment and rebound after treatment cessation (**Figure 3**). After evidence of correlations (via Pearson's test) between individual effects was observed, those correlations were investigated using the full posterior plot of individual effects. Although several potential correlations were discovered, the slopes of these correlations were nearly zero and they were likely detected in error *(type 1)* as an effect of working with such a large dataset (increased statistical power). An even spread of observations vs. individual predictions suggests that this model has no major structural misspecifications and that the error model was well specified (**Figure 4**). However, formal tests for *residual* normality and centering on zero failed. This is likely because of the use of the initial measured tumor volume as the predicted tumor volume at time 0 (i.e. residual is equal to 0) and the conservative estimate of a lower limit of detection of 1 mm for measuring tumor diameter. Once removing the points below the limit of detection and the points measured at time 0, the residual error model aligned much more closely with the theoretical model. The spread of individual parameters met the Kolmogorov-Smirnov test for normality. The precision of parameter estimates was extremely high (RSE < 8%) with low correlation between estimates. Shrinkage of individual parameter distributions toward the mean was low. Full parameter estimates, IIV, and RSEs are reported in **Table 3**. The visual predictive check (VPC) was informative as to wholistic model performance. Although the clinical trials were not matched in terms of sampling schedule, the VPC still indicates overall high-quality fit (**Figure 5**).

*External Validation*

The model was externally validated using the reserved 15% of the data not used during the model building process. Individual parameters were found for each subject, but population parameter estimates, IIV, and error parameters were all fixed to the value found in the final model fit. Model diagnostics were used to assess the ability of the model to make predictions on new patients which were not part of the original dataset. Using the VPC as an overall diagnostic tool (**Figure 6**), the final model fit the external data extremely well with little misspecification except in the lower prediction interval band. This VPC was produced without re-estimating the population parameters, so it is a diagnostic for the fit of the model to an external dataset. Then, the reserved 15% of data was reduced to three subsets, one consisting of only the first datapoint for each patient, another consisting of only the first two datapoints, and a third consisting of only the first three datapoints. Individual predictions for each simulation scenario were compared between the three datasets to assess the quality of the model's individual predictions (**Figure 7**). The model could be used to predict the efficacy of individual treatment time courses with a mean error rate of 59.7% after a single tumor measurement and 11.7% after three successive tumor measurements.

### *Clinical Trial Simulations*

Clinical trial simulations indicated that of the patients who did not go into complete remission in symptoms from standard concomitant therapy, a majority would have seen some percentage in improvement if they had received bevacizumab before pemetrexed-cisplatin. The ideal predicted gap, in steps of 0.1 days, for patients receiving bevacizumab-pemetrexed-cisplatin therapy was 0.4 days or 9.6 hours (**Figure 8**). At this gap, approximately 93.5% of patients benefited from a gap in sequential administration. Of those patients, the mean improvement was 20.7%. The mean improvement in patients at the more practical gaps of 12 and 24 hours were 15.7% and 14.3%, respectively.

## DISCUSSION

The primary objective of this project was to develop a comprehensive semi-mechanistic model that encompasses the growth and response of non-small cell lung cancer (NSCLC) to various clinical antiproliferative treatments. The model produced in this study effectively incorporates the antiproliferative effects of 11 different anti-NSCLC therapeutics used in 11 clinical trials, accounting for both intrinsic and acquired resistance to anticancer therapy. To accomplish this, the most reliable published pharmacokinetic (PK) models for these therapeutics has been integrated into this model, opting not to estimate interindividual variability for PK. The final selected model comprehensively demonstrates transient enhancement of perfusion resulting from anti-VEGF therapy, particularly bevacizumab. The individual predictions derived from this model exhibit a relatively high degree of precision, effectively capturing the well-established rebound in growth that follows the cessation of treatment in non-small cell lung cancer.

While modeling resistance to treatment, AUC was used as a surrogate of drug exposure to model acquired resistance. AUC serves as a meaningful metric for modeling drug exposure, as it is an easily measurable parameter in a clinical setting. It provides a comprehensive representation of the concentration-time profile of a drug, taking into account both the peak concentration and the duration of exposure. By quantifying the total drug exposure over a specified period, AUC enables comparisons across different dosing regimens and facilitates the evaluation of drug efficacy and safety. Therefore, incorporating AUC into the modeling framework allowed for a robust characterization of drug exposure and its relationship to clinical outcomes. Intrinsic resistance has been folded into the distribution of $w_d$ (i.e., weighting) terms in this system of equations. A known weakness in this approach to modeling resistance is that, while capturing a distribution of resistance time courses with respect to exposure, this model has not accounted for the individual systems and pathophysiology that might better explain and characterize that resistance (33).

During this investigation, it was discovered that incorporating a second cytotoxic effect to account for reversible cellular injury proved essential in capturing both the direct effects of bevacizumab and the impact of medications known to induce reversible cell injury, such as paclitaxel and docetaxel. By employing a predominantly linear combination of differential equation terms, this model offers a seamless avenue for integrating additional drug candidates and further system-level details. Moreover, a set of parameter estimates has been meticulously derived and rigorously validated, and the model code has been made openly accessible *(see data sharing statement)* for future research endeavors focused on non-small cell lung cancer. This transparency and availability aim to facilitate scientific progress and enhance the understanding of NSCLC dynamics and therapeutic interventions.

When evaluating the predictive performance of the model, strong evidence was observed in support of model structure choice. Parameter estimates and individual predictions were made with high precision, which is likely due to the largeness of the dataset included. Model structure was based on biological mechanisms and interpretation of parameters is relatively straightforward – e.g. $\lambda$ parameters define the rate of acquired resistance vs. exposure. Mechanisms grounded in biology afford the model additional longevity. On parameter estimate interpretation, relatively simple naming heuristics were used to aid in interpretation. $w_d$ parameters weight drug action against the tumor i.e., the larger the $w_d$ parameter, the larger the action the drug takes proportionally to both the tumor size and concentration of the drug in plasma. The parameter $p$ indicates the proportion of cells in the injured volume which return to unperturbed tumor growth. Unperturbed tumor growth is governed by parameters $\alpha$ (exponential rate of growth of tumor) and $\beta$ (exponential rate of decrease in growth rate due to nutrient supply limitations).

There are several weaknesses which need to be acknowledged in this approach and addressed in future studies. Large amounts of data on cutting-edge first-line therapies–i.e. pembrolizumab and other immune-checkpoint inhibitors–were not included at this time. There are also several covariates, such as weight, body surface area, gender, and concomitant disease which have not been included or tested in the model. Lastly, this modeling project was semi-mechanistic in its aims. This is partially due to the nature of measurements of tumor size made in a clinical trial. Those measurements are sparsely made and strongly influenced by the technicians performing tumor measurements. There was a richness in sampling *between* clinical trials, but within any patient sampling was infrequent. This resulted in imprecise modeling heuristics such as modeling resistance as a function of AUC. However, the platform produced in this study is heavily informed by clinical practice and will make an impactful platform for further iteration informed by either more mechanistic mechanisms or AI modeling methods (34–37).

One of the primary features that this model was designed to capture is the transient enhancement of drug delivery after bevacizumab administration. Theoretically, this transient enhancement drives the synergism between bevacizumab and other antiproliferatives. Potentially some other biological phenomenon may drive the synergism between bevacizumab and other antiproliferatives, but this model is ultimately agnostic about the precise biological mechanism of synergism. A natural conclusion, and a finding supported by previously published clinical papers (19,38–42), is that administering bevacizumab before other antiproliferatives should result in the greatest reduction in tumor size. A delay of 9.6 hours between pemetrexed-cisplatin and bevacizumab resulted in the greatest benefit to the virtual patients. At this gap, approximately 93.5% of patients benefited from a gap in sequential administration. Of those patients, the mean improvement was 20.7%. The mean improvement in patients at the more practical gaps of 12 and 24 hours were 15.7% and 14.3%, respectively. This is within a factor of 3 to the prediction of optimal gap made in Schneider, et al. 2019 (1.2 days) using experimental data from xenograft mice (20). In the future, this model could be adapted to meet standard clinical endpoints of late-stage NSCLC clinical trials such as survival (OS) and progression-free survival (PFS) with RECIST criteria. (43).

## Conclusion and Future Directions

This research provides a simulation platform for future *in silico* studies of optimal scheduling of various therapeutics. The individual Bayesian predictions made in the reserved 15% of data showed that the model had little overfitting and can be extended to external datasets. Mean errors in individual predictions with 1, 2, or 3 sampled points were extremely low, even for far off successive timepoints. For future clinical therapy, this study reaffirms, in human patients, the advantage of staggering bevacizumab, pemetrexed and cisplatin therapy. Lastly, this platform is well supported within the context of this study, but with careful refinement and further validation this model is flexible enough to be adapted to accelerating preclinical predictions in related antiproliferatives. Further research using newer modalities such as immune checkpoint inhibitors and newer methods such as AI-powered modeling will benefit greatly from this significant modeling effort.

## Contributors

PI/Committee, review, oversight: Jonathan P. Mochel and Sebastien Benzekry. All other research was a joint effort with Benjamin K. Schneider as the primary researcher.

## Data Availability Statement

The data that support the findings of this study are available from Vivli Center for Global Clinical Research Data but restrictions apply to the availability of these data, which were used under license for the current study, and so are not publicly available. Some simulated data are however available from the authors upon reasonable request to Benjamin K. Schneider and with permission of Vivli Center for Global Clinical Research Data.

## Declaration of Interests

The authors declare no conflict of interest.

## Funding

Research was funded primarily by Iowa State University as well as a salary supplement via Ceva Sante Animale.

## Acknowledgements

This manuscript is based on research using data from data contributors Roche and Eli Lilly that has been made available through Vivli, Inc. Vivli has not contributed to or approved, and Vivli, Eli Lilly, and Roche, are not in any way responsible for, the contents of this publication.

## FIGURES

**Figure 1 Diagram of the Systems Pharmacology Model**

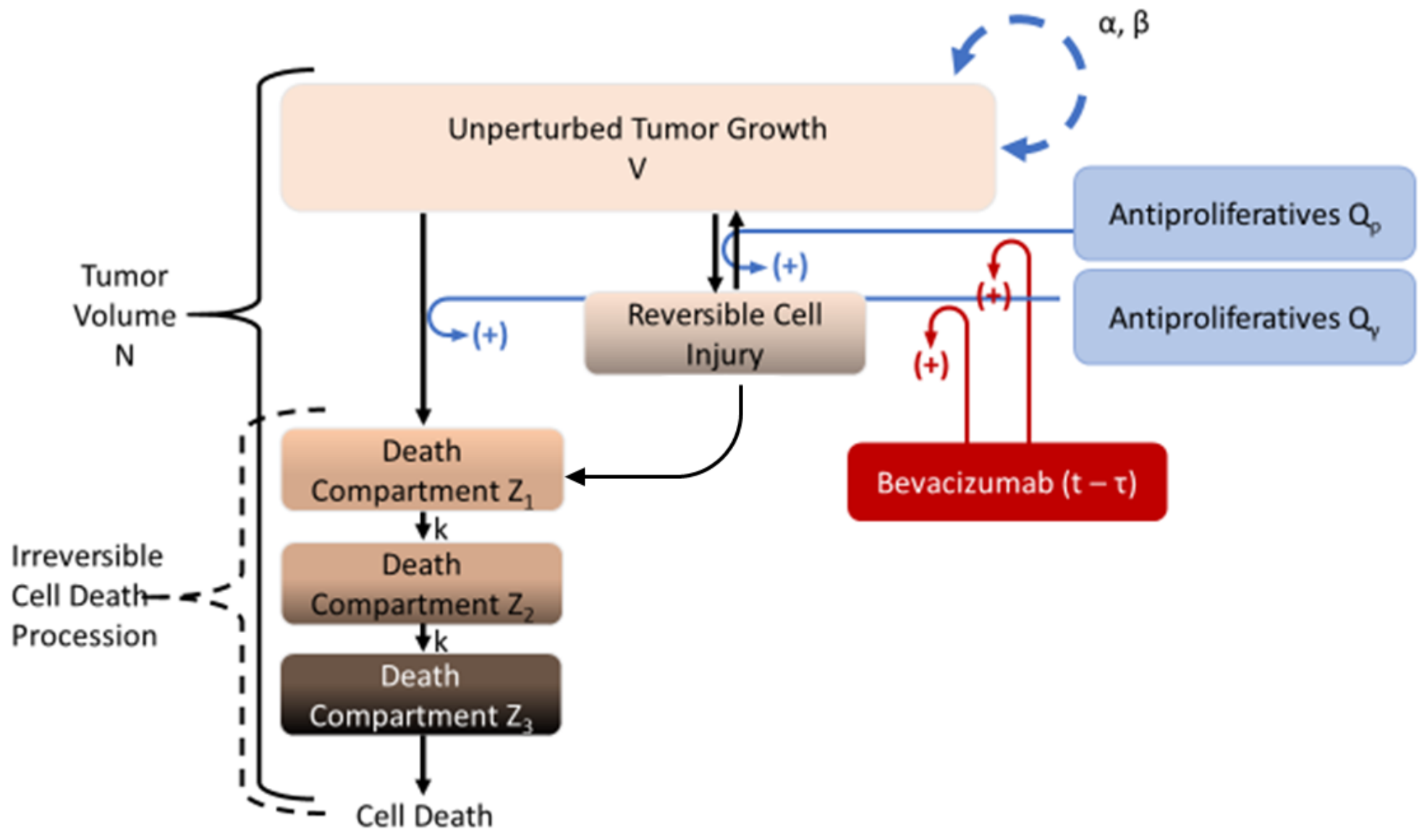

**Comment:** In the Gompertzian model of tumor growth the unperturbed tumor grows at a rate $\alpha$ and is exponentially limited in growth by parameter $\beta$. When a cytotoxic is introduced into the system, the cytotoxic impairs the growth of the tumor by sending cells into a death succession governed by kk. The cytotoxic death relative to drug concentration is modeled as $Q_\gamma$. Reversible cell injury relative to drug concentration is modeled as $Q_\rho$. τ is the time delay between bevacizumab administration and perfusion enhancement effect in the tumor. When a cell is damaged by cytotoxics or reversible cell injury that progresses to irreversible cell injury, it begins a progression from unperturbed growth—compartment V(t)—to damage compartments Z1 through Z1. Eventually the cell exits the tumor volume as it dies.

**Figure 2 SAEM Search**

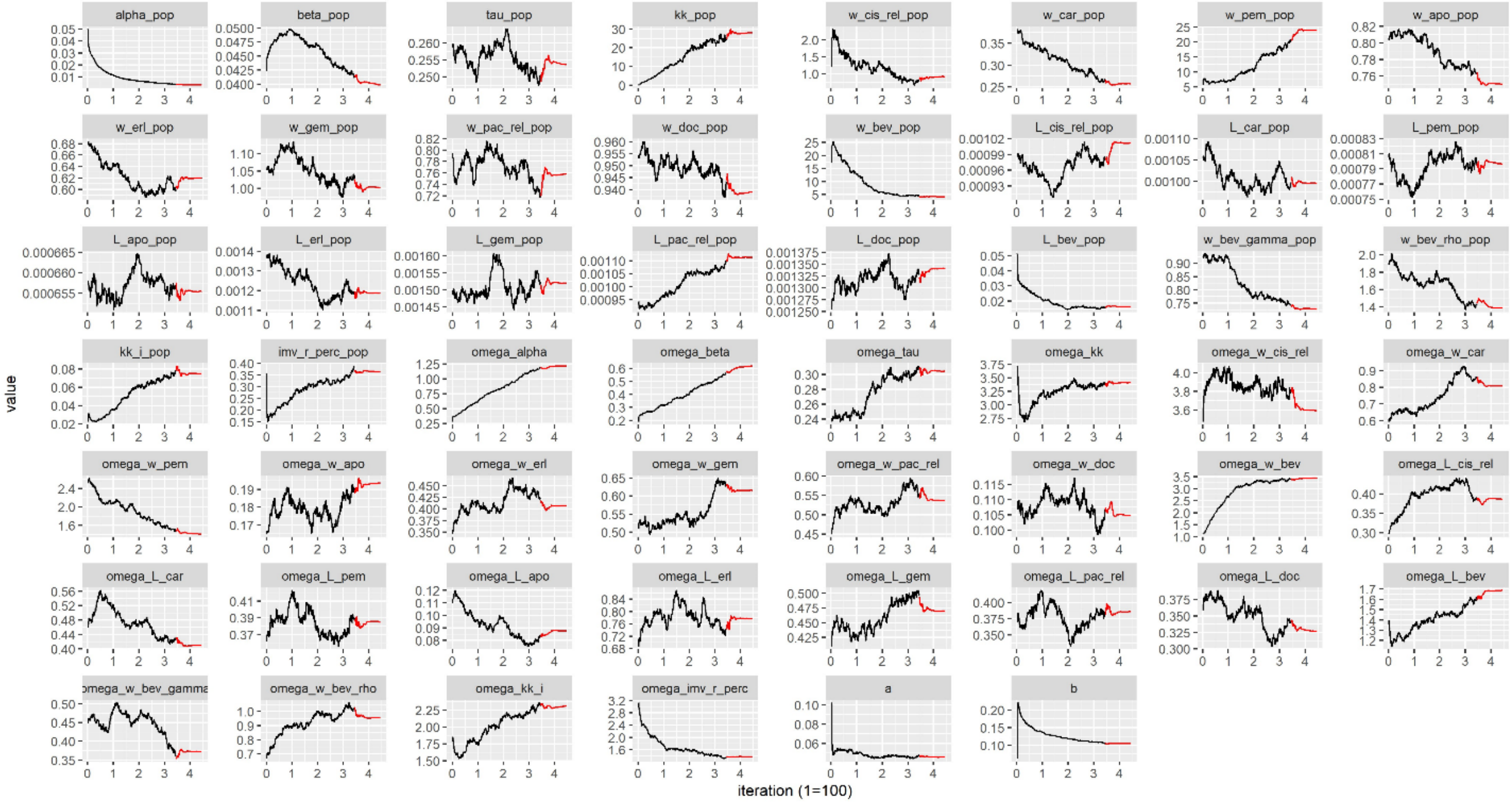

**Comment:** Stochastic approximation expectation maximization search for most likely estimates of parameter values. Exploratory search in black and smoothing search in red. Omega stands for the standard deviation of the random effects.

**Figure 3 Sample of Individual Model Fits**

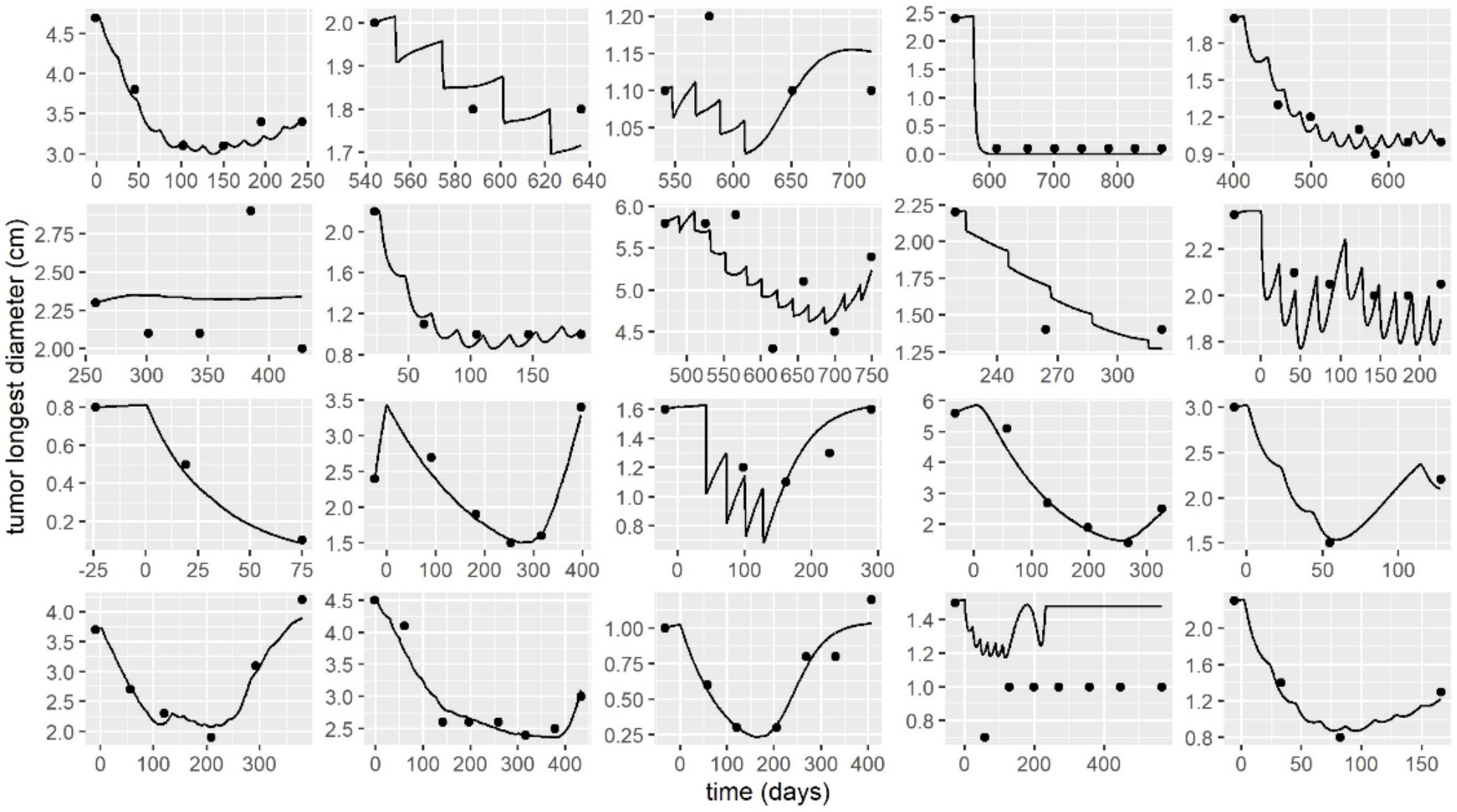

**Comment:** Several individual fits (black continuous line) vs observations (black dots). There's some evidence of overfitting, but the model reproduces a broad range of behaviors over long study periods.

**Figure 4 Observations vs. Predictions Goodness-of-Fits**

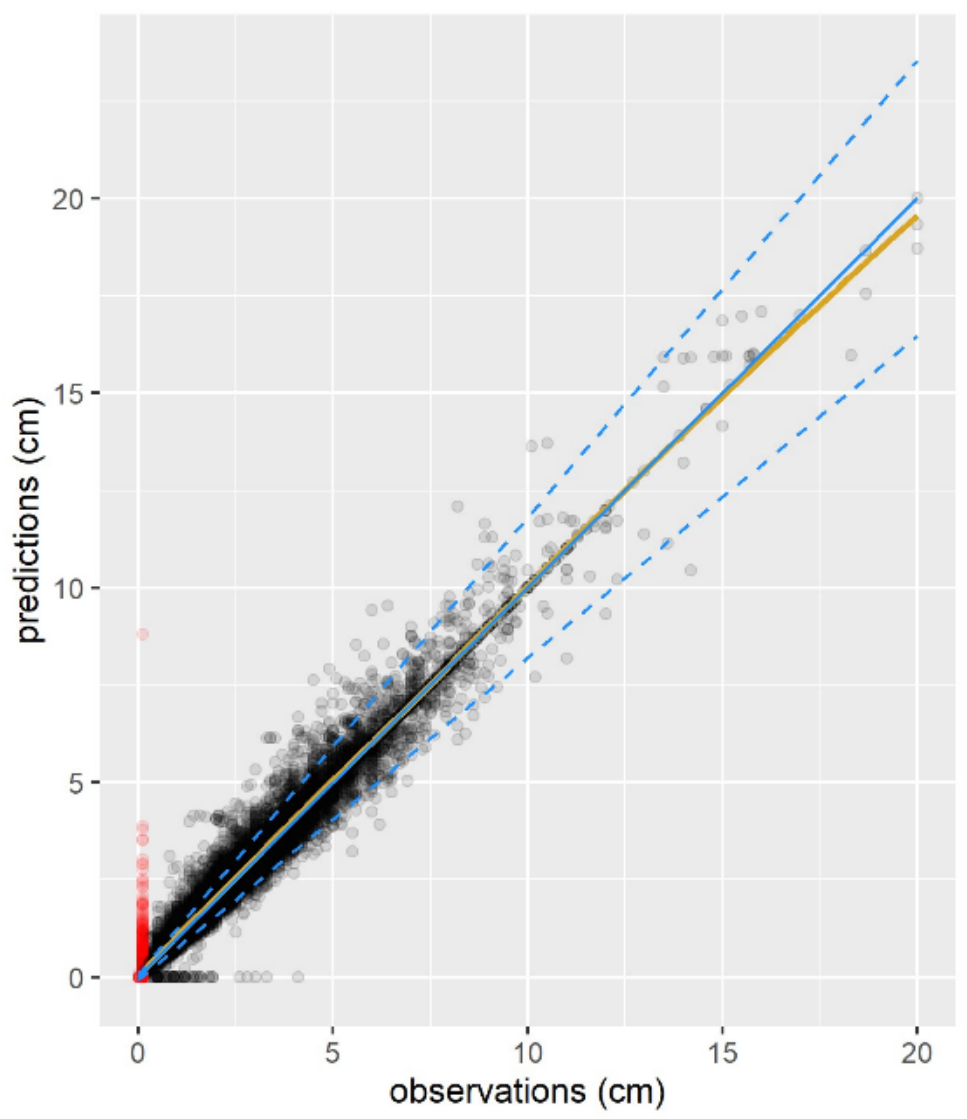

**Comment:** In this figure the observations vs predictions (black points) are plotted along with censored data (red points). Solid line is where observations meet predictions i.e. ratio is 1. Error model 95% prediction boundaries at dotted blue lines. Spline is in yellow. Points are semitransparent to reduce visual overcrowding of data. Relative agreement between spline and line of best fit indicates the overall precision of fit.

**Figure 5 Visual Predictive Check for the Model Fit Using 85% of the Full Dataset (Internal Validation)**

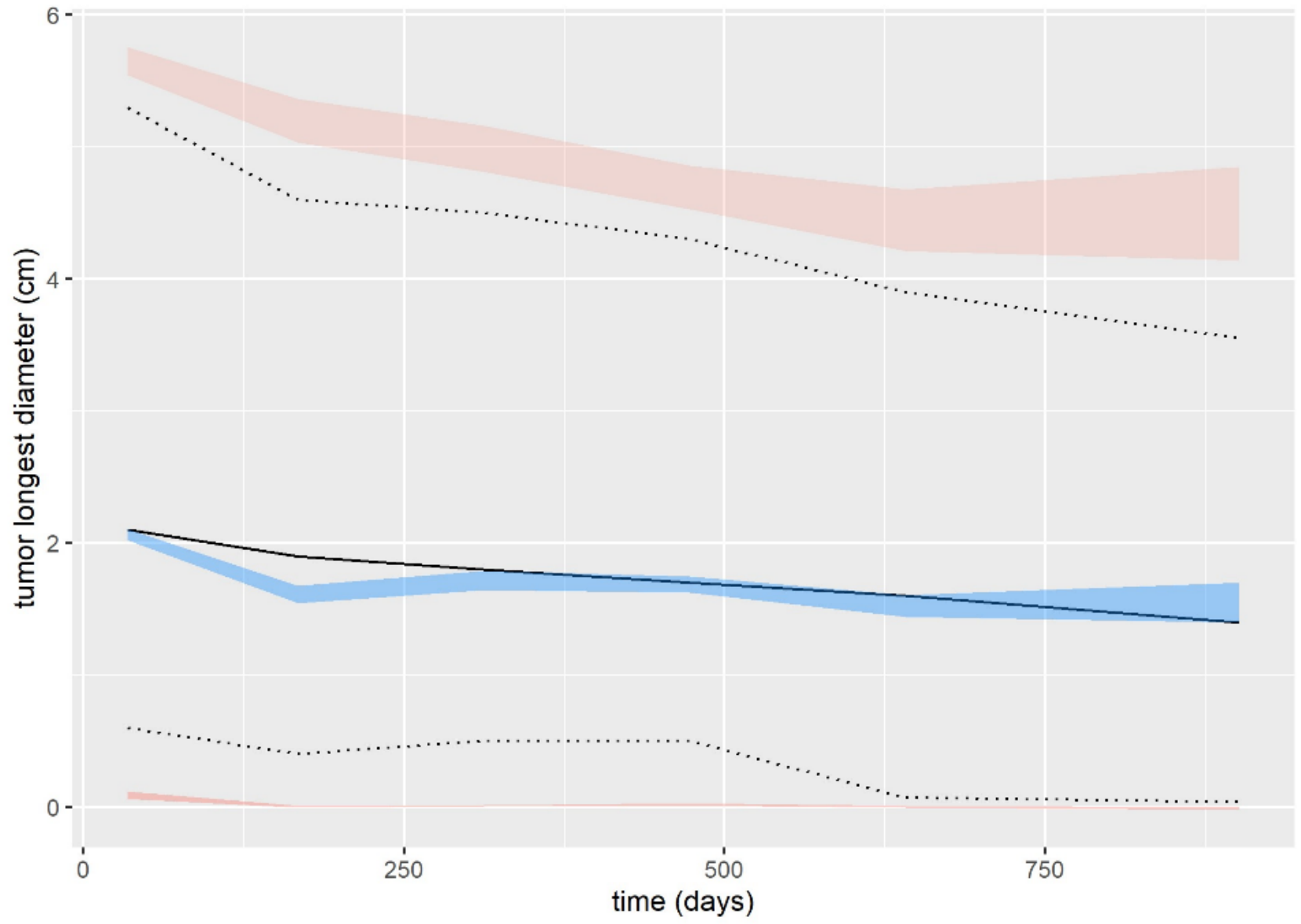

**Comment:** The empirical median tumor longest diameter (central blue line) and upper and lower empirical 90% quantiles (upper and lower blue lines) are predicted precisely by the model fit prediction intervals. The 90% prediction intervals for the median tumor longest diameter (red band) and upper and lower 90% quantiles (blue bands) capture the majority of the variation in the dataset.

**Figure 6 Visual Predictive Check Using 15% of the Full Dataset (Reserved Set: External Validation)**

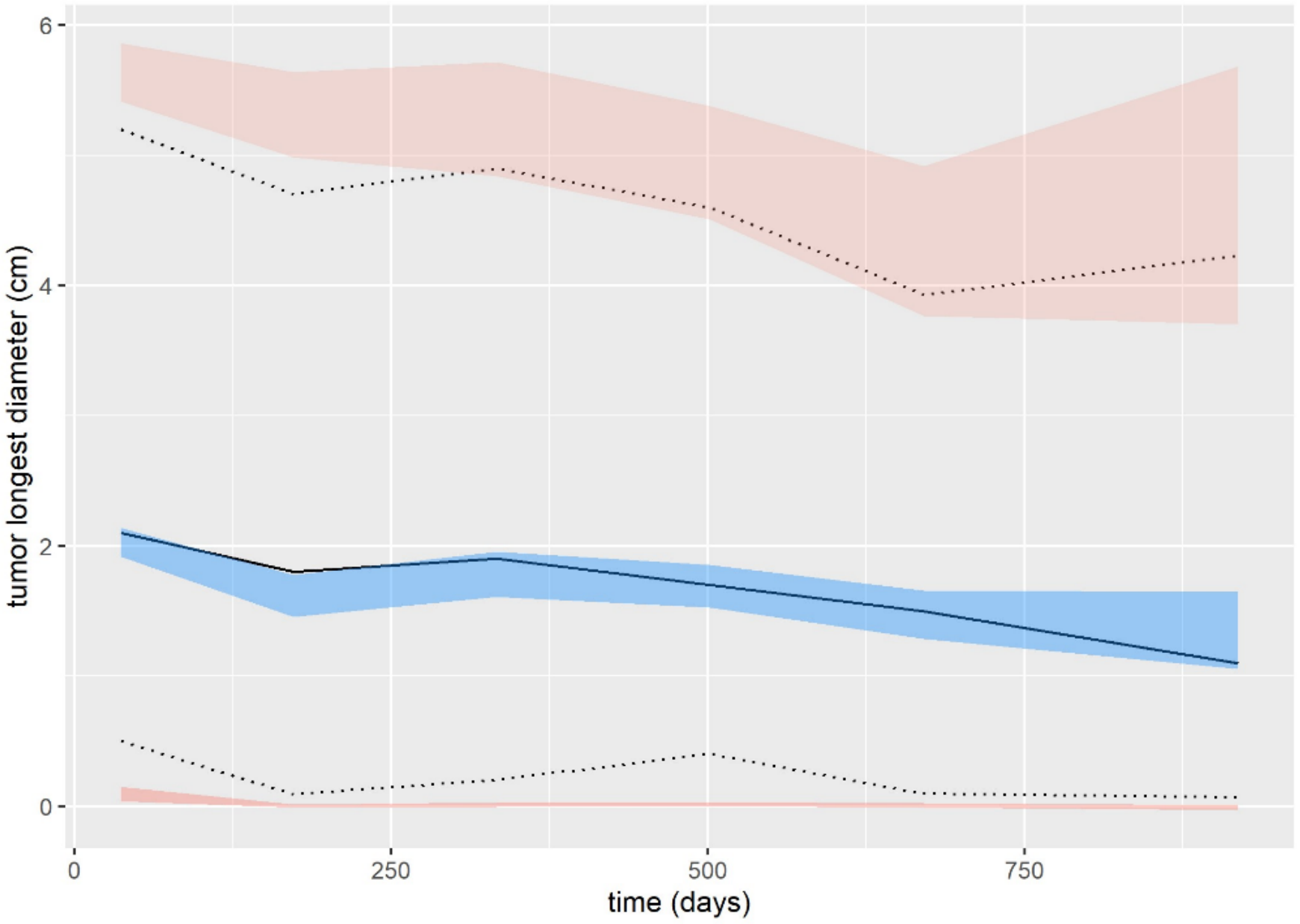

**Comment:** The empirical median tumor longest diameter (central blue line) and upper and lower empirical 90% quantiles (upper and lower blue lines) are predicted precisely by the model fit prediction intervals. The 90% prediction intervals for the median tumor longest diameter (red band) and upper and lower 90% quantiles (blue bands) capture the majority of the variation in the dataset.

**Figure 7 Individual Bayesian Predictions After Limited Sampling**

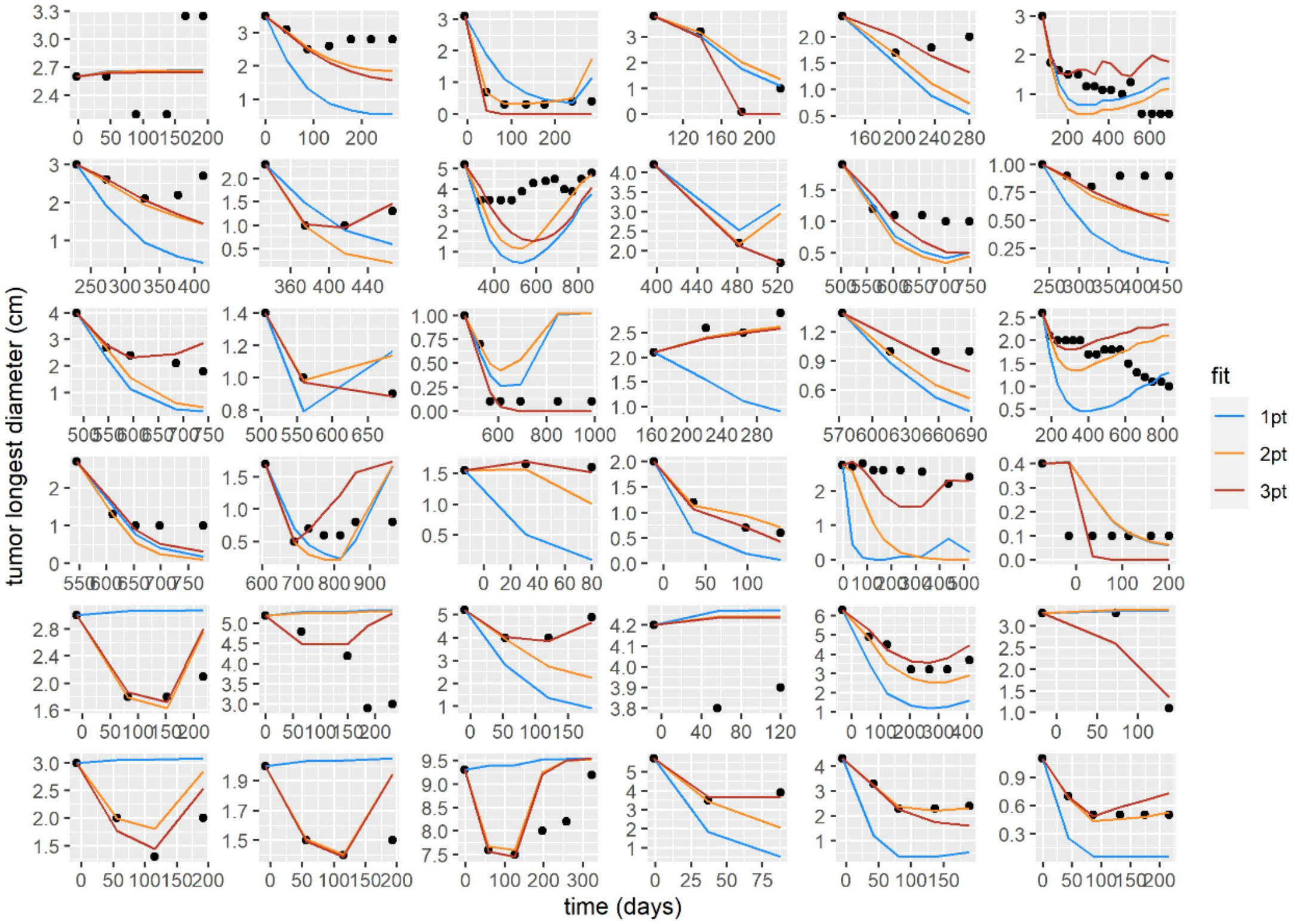

**Comment:** This test was used to assess overfitting in the model. The reserved 15% of data was reduced to three subsets, one consisting of only the first datapoint for each patient, another consisting of only the first two timepoints, and a third consisting of only the first three timepoints. Individual predictions for each were compared between the three datasets to assess the quality of the model's individual predictions. The model could be used to predict the efficacy of individual treatment time courses with a mean error rate of 59.7% after a single measurement and 11.7% after three successive measurements.

**Figure 8 Impact of Gap in Scheduling on Overall Efficacy**

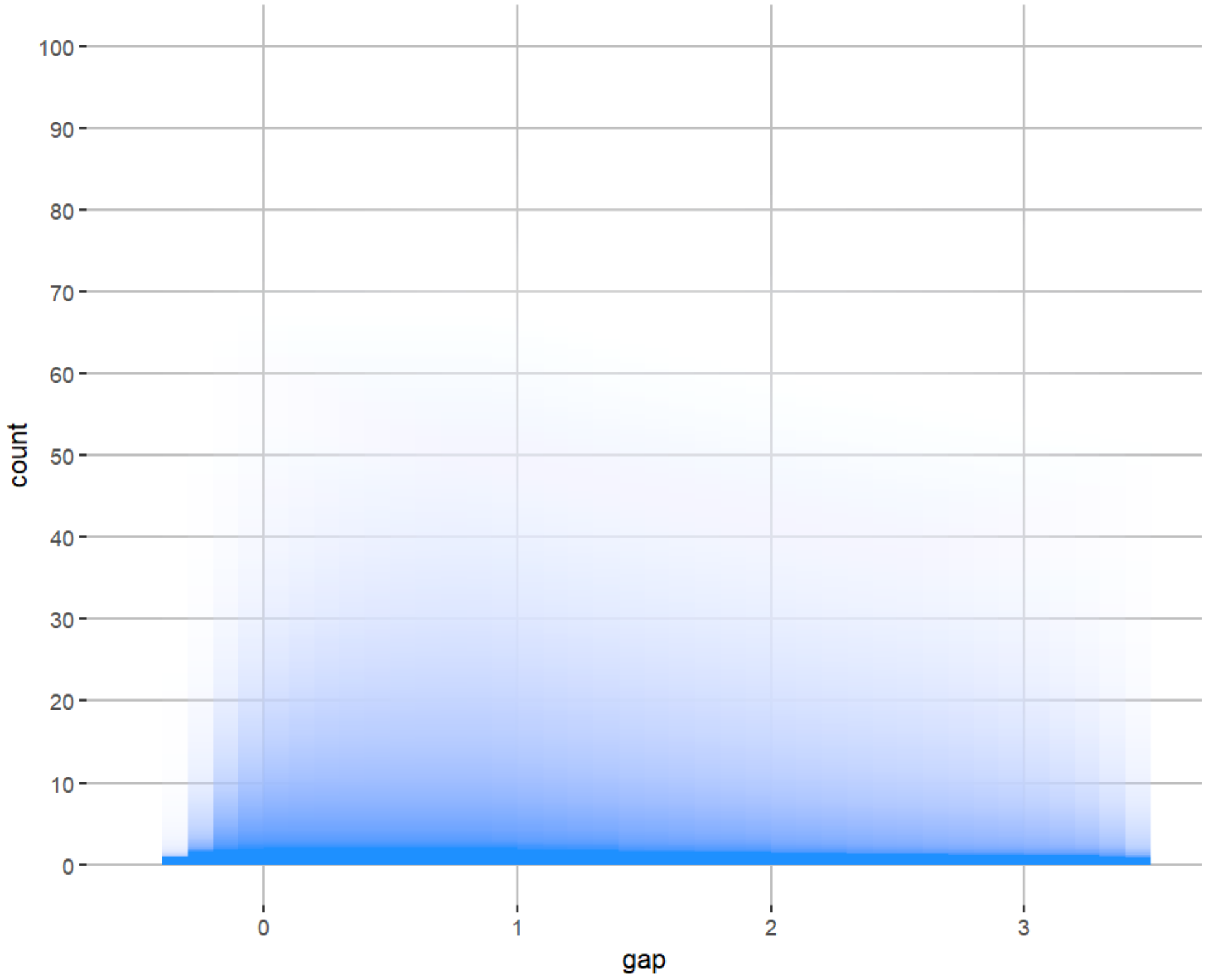

**Comment:** This plot is a combination of a histogram and heatmap for determining the gap where simulated patients saw improvement from delayed bevacizumab treatment. The blue color is the intensity of improvement over concomitant dosing for the simulated patient population at the specified gap in days. The highest intensity of improvement is at a gap of 9.6 hours, or 0.4 days. On the lefthand side, the axis represents percent of patients. After approximately 5 days, there's no longer any improvement gained from delaying bevacizumab administration.

# TABLES

**Table 1 Clinical Data Details**

| CSDR & Vivli | | | |
|---|---|---|---|
| **Trial Registry ID** | Phase | N | Therapeutics |
| NCT00800202 | 2 | 91 | bevacizumab, carboplatin, erlotinib, paclitaxel |
| AVF0757G | 2 | 99 | bevacizumab, carboplatin, paclitaxel |
| NCT01004250 | 2 | 109 | bevacizumab, cisplatin, pemetrexed |
| NCT00312728 | 2 | 115 | bevacizumab, carboplatin, cisplatin, docetaxel, erlotinib, gemcitabine, paclitaxel, pemetrexed, vinorelbine |
| NCT00480831 | 2 | 128 | bevacizumab, carboplatin, paclitaxel, PRO95780 (apomab) |
| NCT00700180 | 2 | 303 | bevacizumab, carboplatin |
| NCT01364012 | 3 | 276 | bevacizumab, carboplatin, paclitaxel |
| NCT00762034 | 3 | 939 | bevacizumab, carboplatin, paclitaxel, pemetrexed |
| NCT00806923 | 3 | 1044 | bevacizumab, cisplatin, gemcitabine |
| NCT02596958 | 4 | 996 | bevacizumab, various platinum-based chemotherapeutics |
| NCT00388206 | 4 | 3998 | bevacizumab, various chemotherapeutics |
| **Additional Vivli Studies Not Available in Time for Modeling** | | | |
| **Trial Registry ID** | Phase | N | Therapeutics |
| doi:10.7150/jca.37966 | N/A | 55 | pembrolizumab, bevacizumab, nivolumab, various chemotherapeutics |
| NCT01846416 | 2 | 138 | atezolizumab |
| NCT01903993 | 2 | 287 | atezolizumab, docetaxel |
| NCT01984242 | 2 | 305 | atezolizumab, bevacizumab, sunitinib |
| NCT02031458 | 2 | 667 | atezolizumab |
| NCT02008227 | 3 | 1225 | atezolizumab, docetaxel |

**Comment:** Data was originally obtained from clinicalstudydataresearch.com, but later during the project Vivli became the custodian of the datasets. During this transition, six more clinical trial datasets representing immune checkpoint inhibitors were added to the repository, but data were not able to be made anonymized and available in time for the modeling project. By following the links to the readers can learn further information about dataset demographics.

**Table 2 Pharmacokinetic Parameters Estimates.**

| *Therapeutic* | *V1* | *V2* | *V3* | *Q1* | *Q2* | *k12* | *k21* | *Cl* | *Source* |
|---|---|---|---|---|---|---|---|---|---|
| *Units* | L | L | L | L/day | L/day | $day^{-1}$ | $day^{-1}$ | L/day | - |
| *bevacizumab* | 2.80 | - | - | - | - | 0.223 | 0.215 | 0.216 | (44) |
| *cisplatin* | 22.3 | 77.0 | - | 456.0 | - | - | - | 6.48 | (45) |
| *pemetrexed* | 12.9 | 3.38 | - | 20.7 | - | - | - | 131.9 | (46) |
| *apomab** | 3.97 | 3.84 | - | .793 | - | - | - | 0.328 | (47) |
| *paclitaxel* | 229.0 | 856.0 | 30.3 | 3216.0 | 5112.0 | - | - | 10296.0 | (48) |
| *carboplatin* | 11.9 | 8.23 | - | 2172.0 | - | - | - | 177.12 | (49) |
| *gemcitabine* | 15.0 | 15.0 | - | 1008.0 | - | - | - | 3888.0 | (50) |
| *docetaxel* | 7.9 | - | - | - | - | 27.12 | 3.6 | 723.12 | (51) |
| *erlotinib*† | 210.0 | - | - | - | - | - | - | 102.96 | (52) |

*see supplementary methods 1; †bioavailability estimated at 60%, ka estimated at 21.36 $day^{-1}$ (52,53)

**Comment: V1** represents volume 1; **V2** represents volume 2; **V3** represents volume 3; **Q1** represents intercompartmental clearance between volumes 1 and 2; **Q2** represents intercompartmental clearance between volumes 2 and 3; **k12** represents rate of transfer from volume 1 to volume 2; **k21** represents rate of transfer from volume 2 to volume 1; **Cl** represents clearance from V1.

**Table 3 Pharmacodynamic Parameters Estimates from the Final Model**

| Fixed Effects | Value | SE | RSE (%) | Unit | IIV |
|---|---|---|---|---|---|
| $\alpha$ | 0.00304 | 8.27E-05 | 2.72 | $day^{-1}$ | 1.22 |
| $\beta$ | 0.0398 | 0.000546 | 1.37 | $day^{-1}$ | 0.62 |
| $\tau$ | 0.254 | 0.00183 | 0.723 | $day^{-1}$ | 0.304 |
| kk | 28 | 2.05 | 7.3 | $day^{-1}$ | 3.41 |
| $w_{cisplatin\ relative\ to\ carboplatin}$ | 0.917 | 0.0728 | 7.94 | - | 3.59 |
| $w_{carboplatin}$ | 0.257 | 0.00486 | 1.89 | - | 0.808 |
| $w_{pemetrexed}$ | 23.8 | 0.759 | 3.19 | - | 1.41 |
| $w_{apomab}$ | 0.75 | 0.00342 | 0.456 | - | 0.194 |
| $w_{erlotinib}$ | 0.62 | 0.006 | 0.968 | - | 0.407 |
| $w_{gemcitabine}$ | 1 | 0.0146 | 1.45 | - | 0.616 |
| $w_{paclitaxel\ relative\ to\ docetaxel}$ | 0.758 | 0.00956 | 1.26 | - | 0.537 |
| $w_{docetaxel}$ | 0.939 | 0.00233 | 0.248 | - | 0.105 |
| $w_{bevacizumab,\ \delta}$ | 4.14 | 0.283 | 6.82 | - | 3.44 |
| $\lambda_{cisplatin\ relative\ to\ carboplatin}$ | 0.00101 | 9.18E-06 | 0.907 | - | 0.386 |
| $\lambda_{carboplatin}$ | 0.000995 | 9.58E-06 | 0.963 | - | 0.409 |
| $\lambda_{pemetrexed}$ | 0.000796 | 7.25E-06 | 0.911 | - | 0.385 |
| $\lambda_{apomab}$ | 0.000656 | 1.35E-06 | 0.206 | - | 0.0877 |
| $\lambda_{erlotinib}$ | 0.00119 | 2.15E-05 | 1.81 | - | 0.777 |
| $\lambda_{gemcitabine}$ | 0.00152 | 1.66E-05 | 1.09 | - | 0.47 |
| $\lambda_{paclitaxel\ relative\ to\ docetaxel}$ | 0.00111 | 1.02E-05 | 0.918 | - | 0.386 |
| $\lambda_{docetaxel}$ | 0.00134 | 1.04E-05 | 0.776 | - | 0.327 |
| $\lambda_{bevacizumab,\ \delta}$ | 0.0164 | 0.000619 | 3.79 | - | 1.69 |
| $w_{bevacizumab,\ \gamma}$ | 0.727 | 0.00638 | 0.878 | - | 0.373 |
| $w_{bevacizumab,\ \rho}$ | 1.39 | 0.031 | 2.23 | - | 0.957 |
| $kk_i$ | 0.0746 | 0.00364 | 4.89 | - | 2.31 |
| ρ | 0.364 | 0.0072 | 1.98 | $day^{-1}$ | 1.37 |
| a | 0.0457 | 0.00116 | 2.53 | - | - |
| b | 0.105 | 0.00105 | 1 | - | - |

**Comment:** SE is the standard error of the estimate and RSE is the relative standard error of the estimate. IIV is the interindividual variability, or in practical terms the standard deviation of the random effects . $\mathbf{w}_{drug}$ parameters are unitless and represent the weight of effect of the treatment modality on Gompertzian growth. $\boldsymbol{\lambda}_{\mathbf{d}}$ parameters are also unitless and represent the effect of AUC on the effectiveness of the relative treatment modalities. **a** and **b** are error model parameters and therefore have no IIV.